\pdfoutput=1
\documentclass{article}
\usepackage[preprint]{neurips_2025}
\usepackage[bottom]{footmisc}
\usepackage[most]{tcolorbox}

\newtcblisting{promptbox}[1]{%
  enhanced, breakable,
  colback=gray!3!white,
  colframe=black!25,
  boxrule=0.5pt,
  arc=3pt,
  left=9pt, right=9pt, top=9pt, bottom=8pt,
  before skip=1.4em, after skip=1.4em,
  fonttitle=\sffamily\bfseries\footnotesize,
  coltitle=black!80,
  colbacktitle=gray!12,
  attach boxed title to top left={yshift=-\tcboxedtitleheight/2, xshift=10pt},
  boxed title style={
    colframe=black!25, boxrule=0.5pt, arc=2pt,
    left=5pt, right=5pt, top=2.5pt, bottom=2.5pt},
  title={#1},
  listing only,
  listing options={
    basicstyle=\ttfamily\footnotesize,
    breaklines=true,
    columns=fullflexible,
    keepspaces=true,
    escapeinside={(*@}{@*)}}
}

\usepackage[utf8]{inputenc}
\usepackage[T1]{fontenc}

\usepackage{amsmath,amssymb,amsfonts}
\usepackage{graphicx}
\usepackage{booktabs}
\usepackage{array}
\usepackage{longtable}
\usepackage{tabularx}
\usepackage{float}

\usepackage{caption}
\usepackage{subcaption}
\usepackage{placeins}

\usepackage{enumitem}
\usepackage{listings}
\usepackage{xcolor}

\usepackage{adjustbox}
\usepackage{pdflscape}
\usepackage{threeparttable}
\usepackage{makecell}

\usepackage{microtype}
\usepackage{seqsplit}

\usepackage{hyperref}

\providecommand{\tightlist}{
\setlength{\itemsep}{0pt}
\setlength{\parskip}{0pt}
}

\hypersetup{
  colorlinks=true,
  linkcolor=blue!55!black,
  citecolor=blue!55!black,
  urlcolor=blue!55!black
}

\title{SureRoute: Toward a Hallucination-Free Self-Improving Platform for Retrosynthesis}
\author{%
Jieli Zhou \quad Naiwu Chen \quad Longzhang Liu \quad Peiyu Zhang* \\
AI Chemistry Group, Future Chemistry Department, XtalPi Inc  \\
\texttt{\{jieli.zhou, naiwu.chen, longzhang.liu, peiyu.zhang\}@xtalpi.com}
}
\begin{document}

\begin{center}
  \colorbox{white}{\includegraphics[height=0.35in]{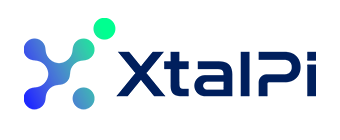}}
\end{center}
\vspace{-1.2em}
\maketitle
\begin{abstract}
AI models, including large language models, are increasingly integrated into
scientific discovery workflows, yet they remain prone to hallucination. In
experimental sciences, such errors translate directly into failed wet-lab
validations and wasted resources; in self-improving agentic systems, confident
errors risk being reinforced rather than corrected. Retrosynthesis provides a
representative example of this failure mode: existing models can generate
chemically plausible routes, but cannot reliably determine which routes are
experimentally feasible. We define \textbf{Chemical Hallucination} as a route
that appears valid yet fails under competing reactive sites, unresolved
selectivity, or missing mechanistic support, a failure largely invisible to the Recall@$K$ metric. We introduce \textbf{SureRoute}, a chemical verifier-anchored
retrosynthesis platform that suppresses Chemical Hallucination. SureRoute
combines a multi-model ensemble, data asset retrieval, and
\textbf{ChemHarness}, an executable chemical intuition engine for route
verification and reliability-first ranking. On a benchmark of 350 real-world
industrial targets, SureRoute reaches 74.3\% recall@1, 2.2--3.5$\times$ that
of seven single-step models and three frontier LLMs, while cutting top-1
Chemical Hallucination to 4.6\%, a 4--6$\times$ reduction relative to frontier
LLMs. As a model-agnostic reranker, ChemHarness drives detectable
hallucination toward near-zero across arbitrary backbone candidates. SureRoute
shows that reliable scientific AI requires not only strong generation, but executable
verification.
\end{abstract}
\section{Introduction}\label{1-introduction}
\vspace{-0.3em}
\noindent\begin{minipage}{\linewidth}
  \centering
  \includegraphics[width=\linewidth,height=2in,keepaspectratio]{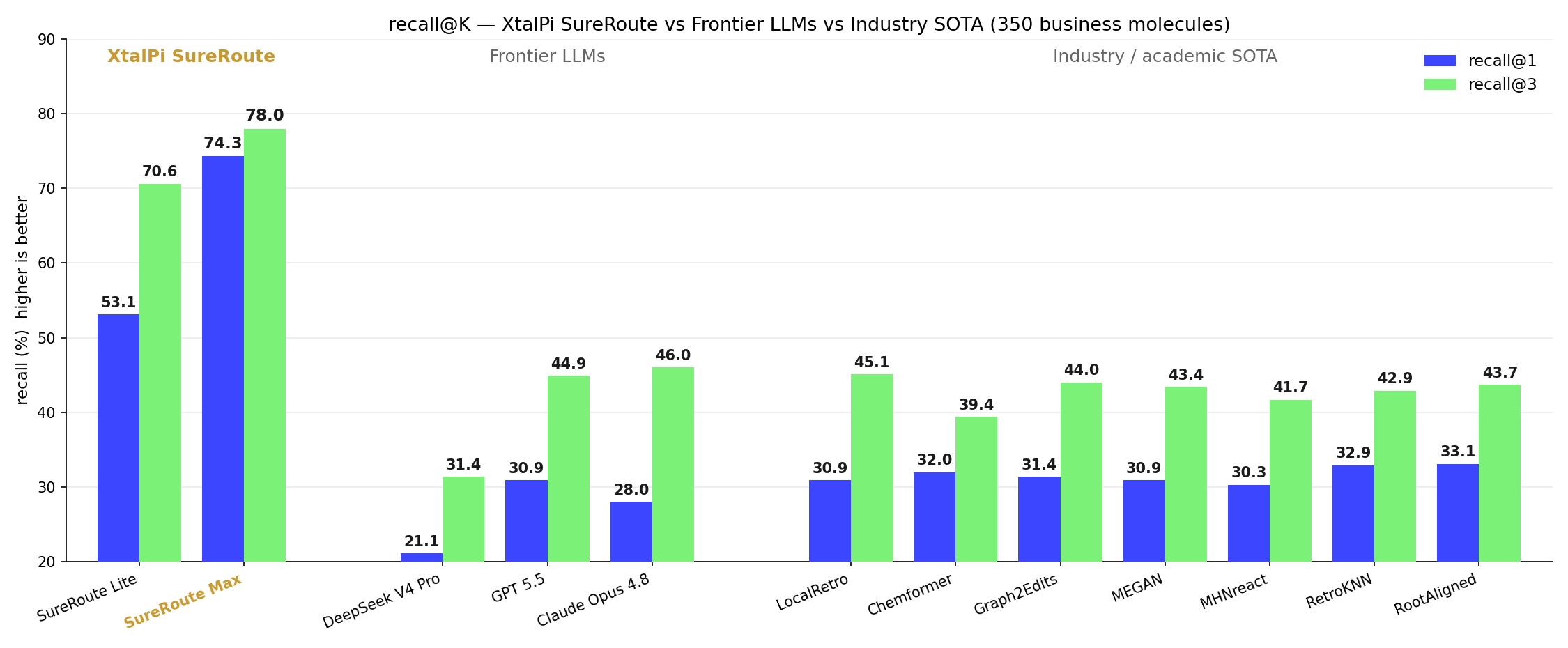}
  \captionof{figure}{recall@1/@3 across 7 single-step models, 3 frontier LLMs, and SureRoute (Lite: ensemble; Max: full system).SureRoute’s recall@1 is a step-function jump above every other method.}
  \label{fig:recall-chart}
\end{minipage}
\vspace{0.4em}

Hallucination---fluent but factually unsupported generation---is a central
obstacle to deploying large language models in high-stakes domains
\citep{ji2023survey,huang2025survey,zhang2023siren}. In dialogue, a
hallucination fact costs a correction; in scientific discovery, it costs an
experiment: an infeasible synthesis route consumes
real laboratory resources before the error is exposed. Models are also poor
judges of their own errors---intrinsic self-correction without external
feedback often fails to improve, and can even degrade reliability
\citep{huang2024large}. The risk is amplified in \emph{self-improving}
systems, where models generate their own training signal through bootstrapped
reasoning, self-refinement, or self-rewarding loops
\citep{zelikman2022star,madaan2023self,yuan2024self}: hallucinations become
training data, and confident errors are reinforced across iterations. What
such loops lack in scientific domains is an \emph{executable verifier}---an
external harness that checks proposals against domain rules rather than the
model's own beliefs.

We study a scientific-discovery setting where this problem is both acute and
measurable: \textbf{retrosynthesis}. Single-step retrosynthesis is the atomic
decision from which every multi-step route is built
\citep{segler2018planning,chen2020retrostar}; if it is wrong, the search tree
expands around a false premise and a plausible-looking route may be mistaken
for an executable plan. Yet for a decade the field has optimized for reference
matching rather than chemical validity: the dominant protocol asks whether
top-$K$ predictions match reference reactants in curated benchmarks such as
the USPTO series \citep{lowe2012extraction,liu2017retrosynthetic,tetko2020state}.
Models optimized for this metric, including recent LLM-based systems, propose
disconnections that are syntactically and superficially plausible---but the
metric never asks whether the transformation would work on \emph{this
substrate}: whether selectivity is controllable, whether competing reactive
sites exist, whether the disconnection has mechanistic or literature precedent.

We call routes that fail on this second axis \textbf{Chemical Hallucinations}:
superficially valid disconnections that an experienced chemist would reject
for unresolved competing reactive sites or lack of mechanistic basis. The
analogy to LLM hallucination is direct---fluent, well-formed, and false---except
the violated constraint is organic chemistry: selectivity, mechanism, and
precedent. This failure is largely invisible to recall@$K$: a model can match
a literature reactant set while ranking chemically dangerous alternatives
highly, and can improve recall without avoiding selectivity traps. In our
evaluation, scaling does not remove this failure mode; frontier LLMs, despite
strong chemical fluency, often produce confident, professionally phrased, and
unexecutable routes.

The stakes extend beyond single predictions. A growing body of work retrains
the single-step policy on the system's own search results, distilling
successful routes back into the model in the spirit of expert iteration and
AlphaGo Zero-style self-play
\citep{anthony2017thinking,silver2017mastering,kim2021self,liu2023pdvn,guo2024resynz}.
This reduces dependence on human-labeled data, but the loop is only as
trustworthy as the \emph{verifier} deciding which self-generated routes to
learn from. Wet-lab validation is trustworthy but slow and expensive; letting
a model score its own output reintroduces exactly the failure this paper
targets. A route that merely \emph{looks} sound can be graded a success and
amplified from an isolated mistake into a systematic prior---a concrete
instance of \emph{misevolution} in self-improving agents
\citep{wang2026misevolve}, made more dangerous here because the error is
chemical rather than syntactic.

This motivates our central question: \textbf{can a verifier approach wet-lab
trustworthiness while retaining the speed and scalability of computational
evaluation?} ChemHarness is our answer: an executable chemical intuition
engine that checks candidate routes for functional-group competition,
regio-/chemoselectivity risk, mechanistic plausibility, and literature
support. SureRoute organizes the entire platform around it as a reliability
layer: generation proposes candidates, verification filters hallucinations,
ranking prioritizes verified routes, and an evolution layer converts
expert-confirmed failures into new ChemHarness data (details in a companion
technical report).

We position SureRoute not as another single-step predictor, but as a
verifier-anchored self-improving platform for reliable retrosynthesis. This
report makes three contributions:

\begin{enumerate}
\item \textbf{SureRoute}, a four-layer platform combining a multi-model
intelligent ensemble, literature-precedent retrieval, the ChemHarness
chemical intuition engine, and an evolution layer that learns from
expert-confirmed failures.
\item \textbf{A reliability-focused evaluation} against seven
state-of-the-art single-step models and three frontier LLMs on 350
real-world industrial molecules, measuring both recall@$K$ and Chemical
Hallucination rate under a unified protocol.
\item \textbf{A demonstration that ChemHarness is model-agnostic and
pluggable}: applied post-hoc to any backbone's top-10 candidates, it drives
detectable top-1 Chemical Hallucination toward near-zero, including for LLMs
never tuned for it.
\end{enumerate}

\section{Related Work}\label{2-related-work}

We organize related work into four threads: single-step retrosynthesis prediction (Section~\ref{21-single-step-retrosynthesis-models}), multi-step planning and search (Section~\ref{22-multi-step-retrosynthesis-planning-and-search}), self-improving retrosynthesis (Section~\ref{23-self-improving-retrosynthesis}), and the broader self-evolution and verification-bottleneck literature in scientific AI (Section~\ref{24-self-evolution-and-the-verification-bottleneck-in-scientific-ai}). The first two threads sketch the specialized technology stack SureRoute builds on; the latter two frame where SureRoute's core contribution sits.

\subsection{Single-step retrosynthesis models}\label{21-single-step-retrosynthesis-models}

Single-step models are commonly grouped by whether they rely on explicit reaction templates. \textbf{Template-based} methods score candidate transformations against a library of reusable rules \citep{segler2017neural,dai2019retrosynthesis}; LocalRetro predicts atom/bond-level local templates on the chemical intuition that reaction changes are mostly local, trading some generalization for interpretability and built-in chemical validity \citep{chen2021localretro}. \textbf{Template-free} methods largely treat retrosynthesis as SMILES-to-SMILES sequence translation---Molecular Transformer-style seq2seq models with SMILES augmentation generalize well but can emit syntactically or chemically invalid output with no built-in feasibility constraint \citep{karpov2019transformer,tetko2020state}. \textbf{Semi-template and graph-based} methods try to combine both strengths: RetroXpert first localizes the reaction center before completing the synthon, and graph-edit formulations represent the transformation as a sequence of graph edits; more recent generative approaches such as Markov-bridge and discrete-flow-matching formulations push single-step diversity and accuracy further \citep{yan2020retroxpert,igashov2024retrobridge,yadav2025retrosynflow}.

For SureRoute, single-step models are a pluggable \textbf{generation layer}: the platform is not tied to any one architecture, but pools multiple single-step models (template-based and template-free alike) behind a unified candidate-proposal interface, leaving organization and filtering to the search and verification layers described in Section~\ref{3-method}.

Beyond dedicated architectures, general-purpose LLMs have recently been applied to retrosynthesis purely via prompting, exhibiting surprisingly strong disconnection recall without task-specific training. LLM retrosynthesis outputs, however, are typically evaluated the same way as specialized models---recall against a reference set---leaving open whether the model's \emph{ranking} of candidates reflects real chemical risk. We treat frontier LLMs as an additional class of backbone under our unified evaluation protocol in Section~\ref{5-experiments}, rather than assuming their fluency implies reliability. 

\subsection{Multi-step retrosynthesis planning and search}\label{22-multi-step-retrosynthesis-planning-and-search}

Above single-step prediction, a multi-step search algorithm determines overall route quality. Monte Carlo Tree Search brought neural-symbolic retrosynthetic planning toward expert-level search, while Retro* proposed AND-OR-tree best-first search guided by a learned value function; later graph-search and experience-guided variants further improved expansion reuse and pruning \citep{segler2018planning,chen2020retrostar,xie2022retrograph,hong2021egmcts}. A notable recent thread is \textbf{planning under uncertainty}: Retro-fallback argues that real-world reaction feasibility is itself stochastic and only partially known, and optimizes for whether a route survives in the lab rather than purely on internal model metrics \citep{tripp2023retrofallback}. Re-evaluation frameworks such as Syntheseus further emphasize the importance of unified, fair multi-step benchmarking, which directly informs our own benchmark design in Section~\ref{4-benchmark-industrial-evaluation-set} \citep{maziarz2023syntheseus}.

\subsection{Self-improving retrosynthesis}\label{23-self-improving-retrosynthesis}

The line of work most directly related to SureRoute feeds search results back into training, closing a self-improvement loop. Its lineage traces to Expert Iteration and AlphaGo Zero: a strong policy is distilled from the outcomes of its own search, then the strengthened policy searches better, and the cycle repeats \citep{anthony2017thinking,silver2017mastering}. Self-Improved Retrosynthetic Planning is an early retrosynthesis instance of this idea \citep{kim2021self}.

PDVN frames retrosynthesis as a tree-structured MDP and trains a policy with dual value networks (synthesizability and cost) directly on whether complete routes solve, rather than only on single-step accuracy \citep{liu2023pdvn}. ReSynZ more directly replicates the AlphaGo Zero recipe for template-based retrosynthesis---MCTS plus reinforcement learning self-improvement---and shows that a strong verifier can substitute for a large training corpus \citep{guo2024resynz}.

The introduction of large language models has pushed this line further. RetroDFM-R drives reasoning-style retrosynthesis with chemically verifiable rewards; Retro-R1 trains an LLM agent to call single-step tools over multiple turns via reinforcement learning; RTRL exploits round-trip consistency between forward and retro tasks to self-reinforce a chemical LLM without labeled data; and end-to-end chain-of-thought approaches internalize multi-step logic entirely inside a single reasoning model, removing the external search heuristic altogether \citep{liu2026retro,zhang2025retrodfmr,zuo2026retrip}.

A clear trend runs through this body of work: reward signals are shifting from human annotation, through model self-evaluation, and converging on \textbf{verifiable rewards}. Several of the most recent LLM-based methods explicitly build ``verifiability'' into their core objective---which is exactly the point made in Section~\ref{1-introduction}: verifier quality is the lever on which the whole loop turns. But the verifiability on offer in this literature is still almost entirely a \textbf{single-step} notion (does the predicted reactant set exactly match a reference, do atoms balance)---a \textbf{route-level, executable, competition-aware} notion of chemical trustworthiness remains largely unaddressed. This is precisely the gap SureRoute and ChemHarness target: verification is elevated from plausibility to executable verification, so that a self-improvement loop consumes a signal close to wet-lab semantics rather than a model's own optimistic self-assessment.

\subsection{Self-evolution and the verification bottleneck in scientific AI}\label{24-self-evolution-and-the-verification-bottleneck-in-scientific-ai}

Zooming out from retrosynthesis to scientific AI broadly, recent surveys organize self-evolving agents into four carrier paths \citep{fang2025survey}: \textbf{weight evolution} (a model updates its own parameters from self-generated data), \textbf{memory evolution} (parameters stay frozen while an external experience store grows), \textbf{tool/code evolution} (the evolving artifact is executable code), and \textbf{physical closed-loop evolution} (wet-lab experiments serve as the ultimate verifier). Self-improving retrosynthesis (Section~\ref{23-self-improving-retrosynthesis}) sits mainly on the weight-evolution path.

On the \textbf{memory-evolution} path, ChemAgent lets an LLM improve with experience via self-updating planning/execution/knowledge memories without any parameter update \citep{tang2025chemagent}. On the \textbf{tool/code-evolution} path, FunSearch and AlphaEvolve show that evolving \emph{programs} rather than answers lets the verifier be cheap and exact (a unit test), while the Darwin G\"odel Machine and G\"odel Agent let an agent rewrite its own code---ChemHarness's practice of compiling chemical knowledge into executable, rule-based checks is the chemistry-domain analogue of this path \citep{romera2023funsearch,novikov2025alphaevolve,zhang2025darwin,yin2025godel}. On the \textbf{physical closed-loop} path, autonomous experimentation systems demonstrate the most trustworthy verifier available---real reaction outcomes---but also the highest iteration cost; a recent survey of self-driving laboratories emphasizes that most deployed systems remain far from full autonomy \citep{boiko2023coscientist,szymanski2023alab,burger2020mobile,lee2026sdl}.

A shared bottleneck runs through all four paths, and it has recently shifted from ``generation capability'' to ``verification capability.'' Work on emergent misevolution risk in self-improving agents makes this shift explicit from a safety angle \citep{wang2026misevolve}: self-evolution without a trustworthy verifier will systematically amplify bias and hallucination rather than correct it. \textbf{SureRoute's position responds to this from the engineering side rather than pushing verification all the way to the most expensive physical extreme}: it occupies a middle ground between weight-evolution loops and physical closed-loop verification---using auditable, executable computational verification (ChemHarness's competition detection and hallucination typing) as a cheap-but-trustworthy reward signal, and using a layered annotation pipeline to accumulate each cycle's failure cases as a compounding asset. Together, these give a retrosynthesis system that can both keep improving itself and resist misevolution while doing so.

\section{Method}\label{3-method}

\begin{figure}[H]\centering
\includegraphics[width=\linewidth]{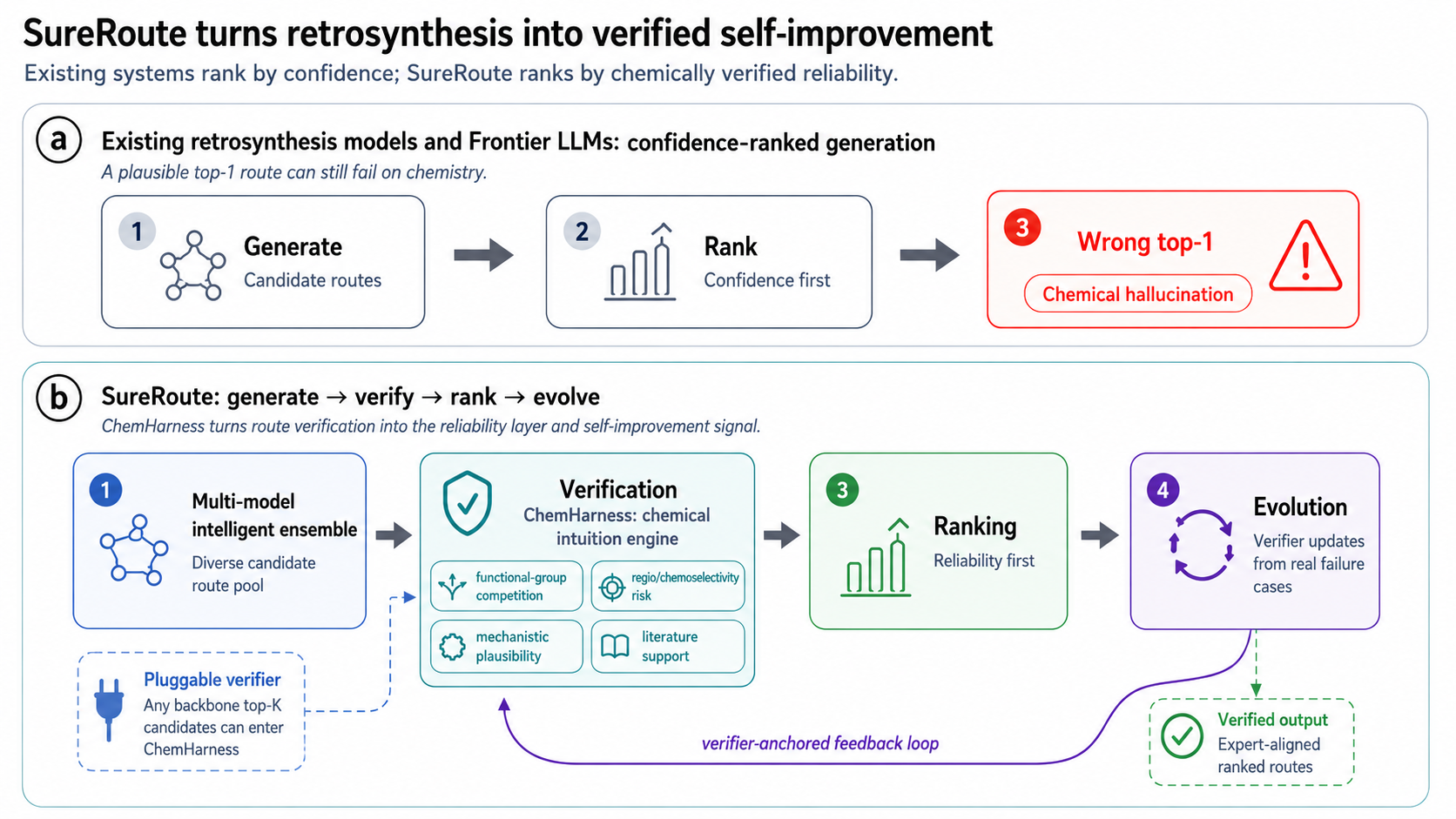}
\caption{\textbf{SureRoute converts retrosynthesis from confidence-ranked generation into verifier-anchored self-improvement.} Existing retrosynthesis systems typically generate candidate routes and rank them by model confidence, which can surface a plausible but chemically hallucination top-1 route. SureRoute instead uses a four-layer architecture: a generation layer builds a diverse candidate pool; ChemHarness verifies each candidate for competition, selectivity, mechanism, and precedent; the ranking layer prioritizes chemically verified routes over confident hallucinations; and the evolution layer converts expert-confirmed failures into high-quality data. The verifier is pluggable: any backbone model's top-$K$ candidates can enter the verification layer.}
\label{fig:sureroute_overview}
\end{figure}

SureRoute is not a single model but a four-layered self-improving system. Each layer addresses a failure mode that the previous layer alone cannot solve (Figure~\ref{fig:sureroute_overview}, Table~\ref{tab:layers}).

None of the four layers is sufficient alone: generation-only ensembles can still walk into selectivity hallucinations; retrieval alone is limited to known precedent; verification without a broad candidate pool has nothing to promote; and self-improvement without a trustworthy verifier risks reinforcing confident mistakes. SureRoute combines all four so the delivered top-1 route is selected by chemically verified reliability rather than raw model confidence.

\begin{table}[H]
\small
\centering
\caption{The four layers of SureRoute and the failure mode each addresses.}
\label{tab:layers}
\vspace{0.4em}
\begin{tabularx}{\linewidth}{p{0.16\linewidth} p{0.30\linewidth} X}
\toprule
Layer & Component & Addresses \\
\midrule
Generation & Multi-model retrosynthesis ensemble & Expands the candidate pool and reduces single-model blind spots across reaction families and scaffolds. \\
Verification & ChemHarness & Screens every candidate for functional-group competition, selectivity risk, mechanistic plausibility, and precedent support. \\
Ranking & Chemical-reliability-aware reranking & Promotes clean, precedent-supported routes above high-confidence hallucinations, changing top-1 selection from model confidence to verified feasibility. \\
Evolution & Expert-confirmed failure feedback loop & Converts rejected production routes into high-quality training data that strengthen the verifier over time. \\
\bottomrule
\end{tabularx}
\end{table}

\subsection{Multi-model ensemble candidate pool}\label{31-multi-model-ensemble-candidate-pool}

For each target molecule, we independently run multiple SOTA single-step retrosynthesis models and collect each model's top-10 ranked candidate disconnections. We develop a novel model ensemble strategy to combine these disconnections. On a high-level, ensemble follows the reciprocal-rank-fusion principle used in information retrieval \citep{cormack2009reciprocal}. Because different architectures (template-based, graph-edit, retrieval-augmented, and transformer-based) fail on different substrates, we pool candidates from all models and aggregate rank information via \textbf{Reciprocal Rank Fusion (RRF)}: each unique (canonicalized) reactant set receives a score

\[s(c) = \sum_{m \in \text{models}} \frac{1}{k_c + \text{rank}_m(c)}\]

where \(\text{rank}_m(c)\) is the rank the candidate received from model \(m\) (candidates not proposed by a model do not contribute), and \(k_c\) is a constant damping the influence of low-rank positions. This produces a single deduplicated, RRF-ranked candidate pool per target that is broader than any single model's coverage---a purely rank-fused ensemble (SureRoute Lite) already greatly improves recall@1 over the strongest single model (Section~\ref{51-disconnection-recallk}), before any reliability layer is applied.

Candidates are compared and deduplicated after a canonicalization step (Section~\ref{4-benchmark-industrial-evaluation-set}) that strips reagents/solvents/salts from both the predicted and reference sides and reduces each candidate to a set of RDKit InChIKeys, so that superficially different SMILES representing the same reactant set are correctly merged \citep{landrum2024rdkit}.

\subsection{ChemHarness: chemical reliability verification}\label{32-chemharness-chemical-reliability-verification}

ChemHarness is treated in this report as the verification layer of SureRoute. The details of ChemHarness will be described in a separate technical report. Here, we focus on the system-level question: given a candidate pool from modern retrosynthesis models and LLMs, can an executable chemical verifier reliably demote hallucination routes and improve top-ranked retrosynthetic suggestions?

At a high level, ChemHarness evaluates each candidate disconnection along different reliability dimensions.

\textbf{Competition and selectivity risk.} ChemHarness checks whether the proposed route contains unresolved functional-group competition or regio-/chemoselectivity ambiguity. These risks arise when a substrate contains multiple chemically plausible reactive sites, and the model proposes a disconnection without explaining why the desired site should dominate. Rather than trusting the model's confidence score, SureRoute uses ChemHarness to identify such selectivity hallucinations and prevent them from being over-ranked.

\textbf{Mechanistic plausibility.} ChemHarness also checks whether the proposed transformation is consistent with the expected chemistry of the substrate and reaction class. This targets a common failure mode of both single-step models and LLMs: proposing a formally balanced disconnection that lacks a viable mechanistic basis. Candidates that fail this check are treated as chemically unreliable even if they appear syntactically valid.

\textbf{Precedent support.} In addition to internal chemical checks, ChemHarness estimates whether a candidate route is supported by relevant literature or database precedent using an intelligent similarity metric. This signal is not used as a blind lookup answer, but as supporting evidence for route reliability: a route with close precedent is favored over a route that is both mechanistically uncertain and unsupported by known chemistry.

The output of ChemHarness is a compact reliability profile for each candidate route, indicating whether the route is likely to be competition-free, mechanistically plausible, and precedent-supported. SureRoute then uses this profile for reliability-first reranking: chemically verified candidates are promoted, while routes that look plausible but contain unresolved chemical risks are demoted.

This design deliberately separates \emph{generation} from \emph{verification}. The generation layer is encouraged to be broad and diverse, while ChemHarness acts as a model-agnostic reliability filter that can be applied to candidates from any backbone model. Because ChemHarness is executable and rule-based at the system level, it can be updated through expert-confirmed failures without retraining the underlying generative models. 

\subsection{Reranking: layered candidate ordering}\label{33-reranking-layered-candidate-ordering}

The final top-$K$ route list served to the user is produced by a layered sort, most-important criterion first:

\begin{enumerate}
\tightlist
\item
  \textbf{Literature/database hit}---if the ensemble's disconnection matches a real literature reaction for this exact target, that route is placed first with maximal confidence, since it is a route a chemist has demonstrably already executed.
\item
  \textbf{No functional-group competition}---among remaining candidates, routes ChemHarness has \emph{not} flagged for selectivity risk are ranked above flagged ones, regardless of the underlying model's raw confidence score.
\item
  \textbf{Precedent score}---among otherwise-equal candidates, higher literature-precedent score is preferred; this is also used to diversify the second and third ranked slots so that routes 2--3 are not simply near-duplicates of route 1.
\item
  \textbf{Ensemble confidence (RRF score)}---the residual tie-breaker.
\end{enumerate}

This ordering choice---competition-free before raw-confidence---is deliberate and is the central mechanism by which SureRoute converts ensemble recall into low-hallucination top-1 recall: a route that is individually the single most-confident prediction from the strongest model, but that ChemHarness flags as sitting on a competing site, is demoted below a less-confident but competition-free alternative.

\subsection{Pluggability}\label{34-pluggability}

ChemHarness's competition/plausibility checks and precedent scoring operate purely on a proposed reaction (reactants + product), independent of which model proposed it. This means the reranking procedure in Section~\ref{33-reranking-layered-candidate-ordering} can be applied to the top candidates from \emph{any} backbone---a different single-step model, or a frontier LLM prompted for retrosynthesis---with no retraining. Section~\ref{53-pluggability-reranking-arbitrary-backbones-with-chemharness} demonstrates this directly: applying ChemHarness reranking to LLM and single-step-model candidate pools drives their top-1 Chemical Hallucination rate toward 0\%, even though ChemHarness's rules were developed independently of any of these backbones.

\subsection{ChemHarness as the anchor for a self-improving loop}\label{35-chemharness-as-the-anchor-for-a-self-improving-rule-loop}

As motivated in Section~\ref{1-introduction}, a self-improvement loop for retrosynthesis is only as trustworthy as its verifier. Wet-lab feedback is the gold standard but is far too slow and expensive to gate every candidate route; letting a model score its own output is fast but reintroduces the exact failure this paper documents, since a confidently wrong route can be graded as a success and distilled back into the system. ChemHarness is designed to sit between these two extremes: its competition and plausibility checks are executable and deterministic (Section~\ref{32-chemharness-chemical-reliability-verification}), so a verdict can be produced and audited in milliseconds rather than weeks, while still being anchored in explicit chemical rules rather than a model's self-assessment.

This property lets ChemHarness close a rule-improvement loop entirely within the computational domain. Every route SureRoute serves in production is a candidate observation: when a chemist overrides or rejects a top-ranked route, or when downstream forward-synthesis checks disagree with a route ChemHarness had marked clean, that disagreement is logged as a confirmed failure case rather than discarded. These confirmed failures are triaged into the same rule families described in Section~\ref{32-chemharness-chemical-reliability-verification} and used to either tighten an existing competition rule (adding a reactive-site pattern the rule previously missed) or carve out a new exception (a substrate pattern the rule previously over-flagged). Because the loop's reward signal is a rule violation checked against a real, chemist-confirmed outcome---not a model's own confidence---a route that merely \emph{looks} mechanistically sound cannot be mistaken for a validated success and cannot be distilled back in as a false positive, which is the specific misevolution risk described in Section~\ref{24-self-evolution-and-the-verification-bottleneck-in-scientific-ai}.

Two properties make this loop compound rather than plateau. First, coverage grows with usage: every production molecule SureRoute processes is a potential source of new failure cases, so the rule set's blind spots shrink as deployment scales, without requiring a proportional increase in wet-lab or manual-annotation cost. Second, the loop is auditable at every step---because rules given by ChemHarness are explicit and human-readable rather than distributed across model weights, a chemist can inspect exactly which rule fired, agree or disagree with it, and correct it directly, which is not possible for a black-box learned reranker. This combination---cheap-but-trustworthy verification, plus a rule representation a domain expert can directly edit---is what makes SureRoute's reliability advantage as compounding: it is not a fixed asset delivered once, but a system property that keeps improving with production usage in a way a static, pretrained model cannot replicate without repeating the same accumulation process.

\section{Benchmark: industrial retrosynthesis evaluation set}\label{4-benchmark-industrial-evaluation-set}

\subsection{Motivation}\label{41-motivation}

Standard retrosynthesis benchmarks mainly evaluate whether a model's top-$K$ predicted reactant sets match a reference reaction in curated public datasets such as USPTO \citep{lowe2012extraction,liu2017retrosynthetic,tetko2020state,maziarz2023syntheseus}. This is useful for measuring disconnection recall, but it is not sufficient for measuring route reliability in industrial settings. In real projects, a route can look reasonable at the reactant-set level while still being unsuitable for execution. To evaluate this gap, we constructed an internal industrial retrosynthesis evaluation set. 

\subsection{Dataset}\label{42-dataset}

The evaluation set contains 350 real industrial molecules selected from practical retrosynthesis use cases. These molecules are more challenging than textbook examples: they contain realistic scaffolds, multiple functional groups, and substrate-specific constraints that require chemical judgment rather than pattern matching alone.

For each target, we collected expert-validated reference routes and compared model-generated candidates against these references under a unified evaluation protocol.

\begin{figure}[H]\centering
\includegraphics[width=0.95\linewidth]{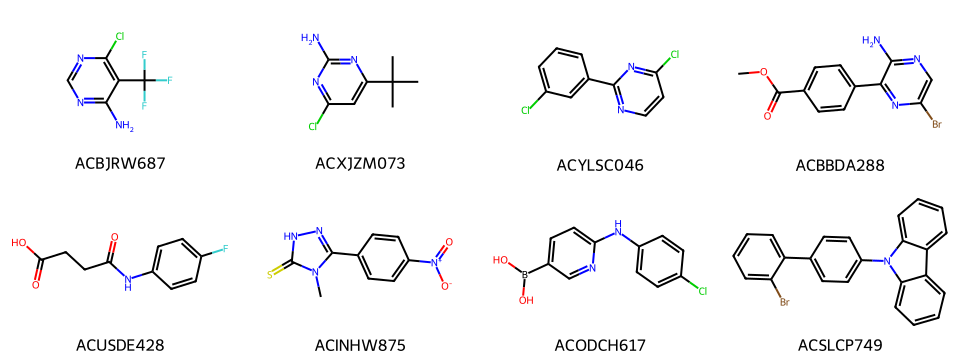}
\caption{\textbf{Representative molecules from the internal industrial evaluation set.} The benchmark focuses on realistic industrial targets rather than simplified textbook reactions. Molecule IDs can be queried from the XtalPi Aifchem website \url{https://www.aifchem.com/}.}
\label{fig:benchmark_gallery}
\end{figure}

\subsection{Evaluation protocol}\label{43-evaluation-protocol}

\textbf{Task.} Given a target molecule, each system proposes up to 5 ranked single-step retrosynthetic candidate reactant sets.

\textbf{Matching.} A prediction is counted as a hit only when the predicted reactive-core reactant set matches the expert-validated reference after the same standardization procedure is applied to all methods. Reagents, solvents, catalysts, salts, and other non-core components are handled consistently across models so that the comparison measures retrosynthetic disconnection quality rather than formatting differences.

\textbf{Chemical reliability.} In addition to recall@$K$, we evaluate whether the top-ranked route contains chemical reliability risks. This allows us to measure not only whether a model can generate a reference-like answer, but also whether its top recommendation is chemically trustworthy. All routes are carefully inspected and annotated by an expert chemist. 

\textbf{Fair comparison.} All single-step models, LLMs, and SureRoute variants are evaluated with the same targets, same candidate limit, same standardization pipeline, and same scoring criteria. Literature-retrieval matches are excluded from the main recall comparison when needed to avoid inflating performance through direct database overlap.

\section{Experiments}\label{5-experiments}

\subsection{Disconnection recall@$K$}\label{51-disconnection-recallk}

\begin{table}[H]
\small
\centering
\caption{recall@$K$ on the 350-molecule benchmark.}
\label{tab:recall}
\vspace{0.4em}
\begin{tabularx}{\linewidth}{X r r r}
\toprule
Method & recall@1 & recall@3 & recall@5 \\
\midrule
LocalRetro & 30.9\% & 45.1\% & 49.1\% \\
Chemformer & 32.0\% & 39.4\% & 40.3\% \\
Graph2Edits & 31.4\% & 44.0\% & 49.4\% \\
MEGAN & 30.9\% & 43.4\% & 45.7\% \\
MHNreact & 30.3\% & 41.7\% & 44.3\% \\
RetroKNN & 32.9\% & 42.9\% & 46.0\% \\
RootAligned & 33.1\% & 43.7\% & 46.3\% \\

DeepSeek V4 Pro & 21.1\% & 31.4\% & 38.3\% \\
GPT-5.5 & 30.9\% & 44.9\% & 50.9\% \\
Claude Opus 4.8 & 28.0\% & 46.0\% & 54.0\% \\
SureRoute Lite (model ensemble, no ChemHarness) & 53.1\% & 70.6\% & 72.3\% \\
\textbf{SureRoute Max (model ensemble, with ChemHarness)} & \textbf{74.3\%} & \textbf{78.0\%} & \textbf{78.6\%} \\
\bottomrule
\end{tabularx}
\end{table}

Table~\ref{tab:recall} and Figure~\ref{fig:recall-chart} report recall@$K$ under the protocol in Section~\ref{43-evaluation-protocol}, for 7 SOTA single-step models, 2 SureRoute configurations, and 3 frontier LLMs. The specialized backbones represent the main template-based, template-free, graph-based, and retrieval/alignment-style families developed in the retrosynthesis literature \citep{segler2017neural,chen2021localretro,karpov2019transformer,tetko2020state,yan2020retroxpert}.

SureRoute Max's recall@1 of 74.3\% is 1.4\ensuremath{\times} the plain-ensemble baseline (SureRoute Lite, 53.1\%) and 2.2--3.5\ensuremath{\times} every single-step model (30--33\%) and every frontier LLM (21--31\%). Because a chemist typically acts on the first proposed route, recall@1 is the most operationally relevant number in this table, and the gap here is far larger than at recall@3 or recall@5, where ensembling alone (without reliability reranking) already closes much of the distance. This is consistent with our thesis: the largest single lever for realistic top-1 recall is not more/better disconnection candidates but \emph{correctly discarding the confident-but-wrong ones}---which is exactly what ChemHarness's reranking (Section~\ref{33-reranking-layered-candidate-ordering}) is designed to do, and which pure ensembling cannot achieve by itself.

\subsection{Chemical Hallucination rate}\label{52-chemical-hallucination-rate}

Recall@$K$ alone cannot distinguish ``the model's top choice is chemically sound'' from ``the model's top choice happens to match the literature answer by chance.'' We therefore report \textbf{Chemical Hallucination rate}: the fraction of targets for which a method's \emph{actual delivered top-1 route} has either functional-group competition or mechanistic implausibility. Lower is better; a route is counted as a hallucination if it fails \emph{either} check. The inspections are carefully done by experienced XtalPi Chemists.

\begin{figure}[H]\centering
\includegraphics[width=0.95\linewidth]{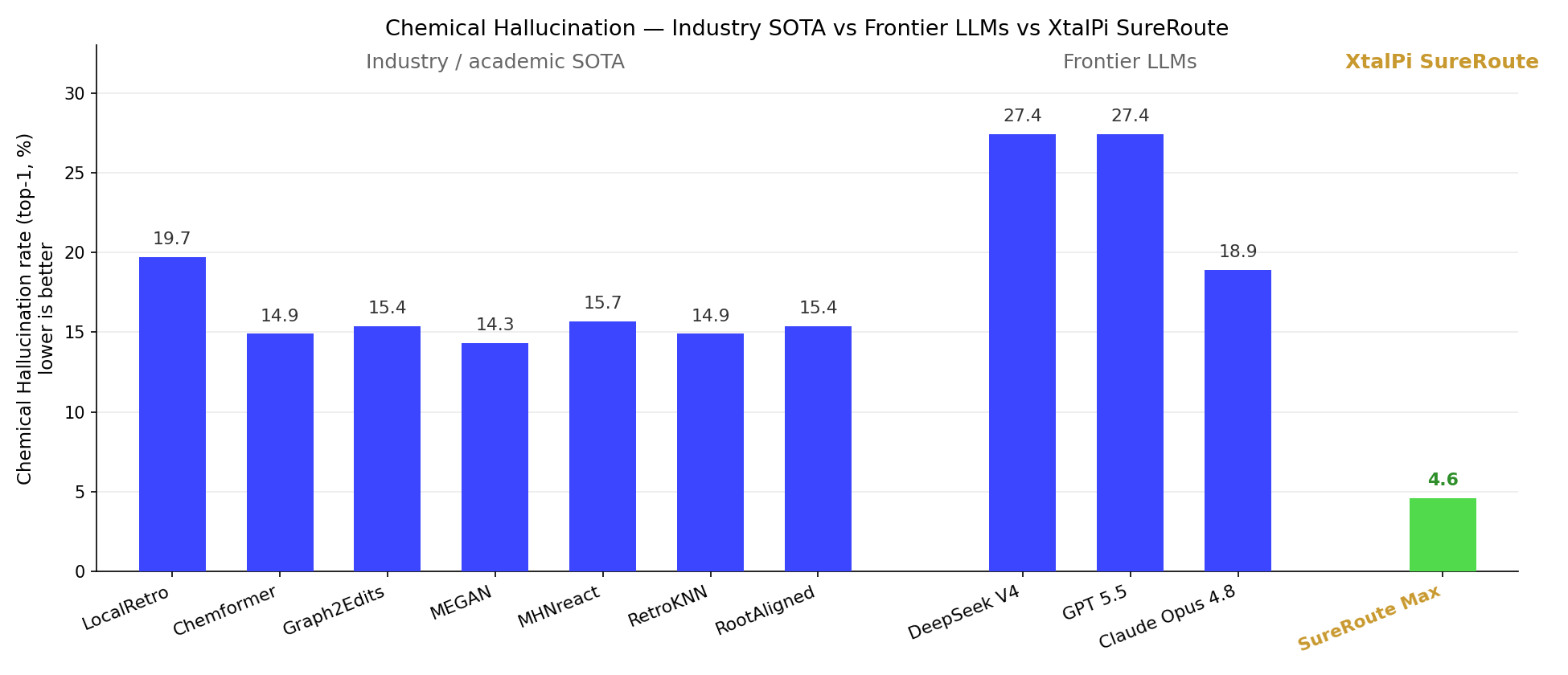}
\caption{Top-1 Chemical Hallucination rate across all methods. Frontier LLMs (18.9--27.4\%) are \emph{worse} than every single-step model (14.3--19.7\%) on this axis despite matching or exceeding them on raw recall (Figure~\ref{fig:recall-chart})---fluent disconnection does not imply selectivity awareness. SureRoute's 4.6\% is the only bar in the single digits.}
\label{fig:hallucination-chart}
\end{figure}

All 7 single-step models hallucinate at 14--20\% on their top-1 route (Figure~\ref{fig:hallucination-chart}, Table~\ref{tab:hallucination}). The three frontier LLMs are \emph{worse}, not better, at 18.9--27.4\%: fluency and disconnection accuracy do not transfer into selectivity awareness. SureRoute's top-1 Chemical Hallucination rate of 4.6\% is achieved by combining literature-route injection with ChemHarness-based reranking (Section~\ref{33-reranking-layered-candidate-ordering}), which actively demotes competition-flagged candidates rather than merely reporting whichever candidate scored highest under the underlying model's own confidence.

\begin{table}[H]
\small
\centering
\caption{Chemical Hallucination rate (top-1 route).}
\label{tab:hallucination}
\vspace{0.4em}
\begin{tabularx}{0.78\linewidth}{X r}
\toprule
Method & Chemical Hallucination (top-1) \\
\midrule
LocalRetro & 19.7\% \\
Chemformer & 14.9\% \\
Graph2Edits & 15.4\% \\
MEGAN & 14.3\% \\
MHNreact & 15.7\% \\
RetroKNN & 14.9\% \\
RootAligned & 15.4\% \\

DeepSeek V4 & 27.4\% \\
GPT-5.5 & 27.4\% \\
Claude Opus 4.8 & 18.9\% \\
\textbf{SureRoute Max} & \textbf{4.6\%} \\
\bottomrule
\end{tabularx}
\end{table}

\subsection{Pluggability: reranking arbitrary backbones with ChemHarness}\label{53-pluggability-reranking-arbitrary-backbones-with-chemharness}

\begin{figure}[H]\centering
\includegraphics[width=0.95\linewidth]{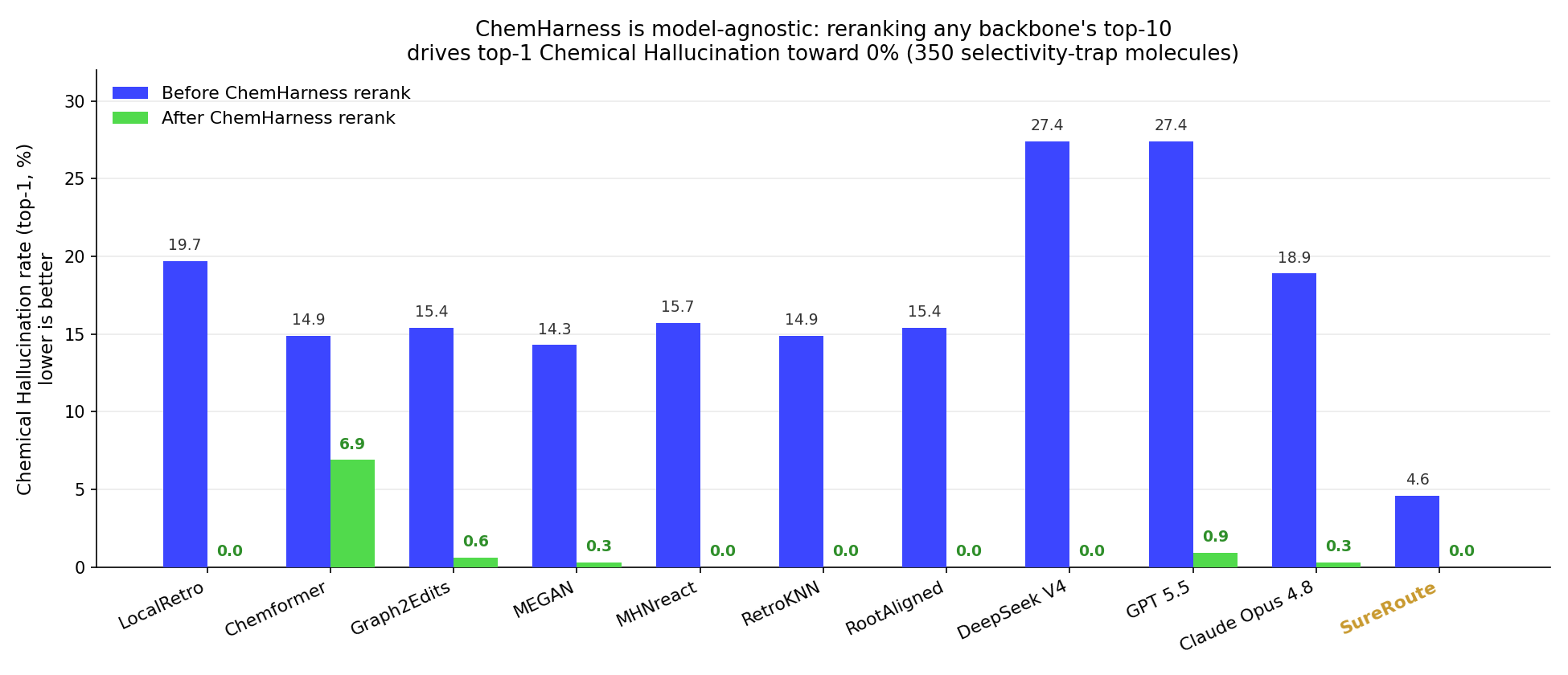}
\caption{Top-1 Chemical Hallucination rate before (blue) and after (green) applying ChemHarness reranking to each backbone's own top-10 candidates, with no other change to the underlying model. Every bar collapses toward zero, including for LLMs ChemHarness was never tuned against.}
\label{fig:pluggability-chart}
\end{figure}

To test whether ChemHarness's benefit is specific to SureRoute's own ensemble or generalizes to arbitrary backbones, we take each method's top-10 raw candidates (single-step models and LLMs alike) and rerank them by ChemHarness's competition/plausibility signal alone (Section~\ref{33-reranking-layered-candidate-ordering}, criterion 2), with no other change to the candidate pool (Figure~\ref{fig:pluggability-chart}, Table~\ref{tab:rerank}).

Reranking collapses ChemHarness-detectable top-1 Chemical Hallucination toward near-zero for every backbone we tested, specialized model and frontier LLM alike, despite ChemHarness's rules having been developed without reference to any of these specific models. This is the central evidence for our pluggability claim: the reliability layer is separable from the disconnection-generation layer, and can be composed with a system it was never tuned against.

\begin{table}[H]
\small
\centering
\caption{Top-1 Chemical Hallucination rate, before $\rightarrow$ after ChemHarness reranking.}
\label{tab:rerank}
\vspace{0.4em}
\begin{tabularx}{0.82\linewidth}{X l}
\toprule
Method & Before $\rightarrow$ After \\
\midrule
LocalRetro & 19.7\% $\rightarrow$ 0.0\% \\
Chemformer & 14.9\% $\rightarrow$ 6.9\% \\
Graph2Edits & 15.4\% $\rightarrow$ 0.6\% \\
MEGAN & 14.3\% $\rightarrow$ 0.3\% \\
MHNreact & 15.7\% $\rightarrow$ 0.0\% \\
RetroKNN & 14.9\% $\rightarrow$ 0.0\% \\
RootAligned & 15.4\% $\rightarrow$ 0.0\% \\
DeepSeek V4 & 27.4\% $\rightarrow$ 0.0\% \\
GPT-5.5 & 27.4\% $\rightarrow$ 0.9\% \\
Claude Opus 4.8 & 18.9\% $\rightarrow$ 0.3\% \\
\textbf{SureRoute Max} & 4.6\% $\rightarrow$ 0.0\% \\
\bottomrule
\end{tabularx}
\end{table}

\textbf{Residual rate as a diagnostic.} Reranking fully eliminates hallucination for candidate-rich models (LocalRetro, MEGAN) but leaves a small residual for Chemformer (6.9\%)---because when a model's entire top-10 candidate pool consists of competition-flagged disconnections, there is no clean alternative for the reranker to promote. The residual rate after reranking is therefore itself informative: it measures whether a backbone's candidate \emph{pool} contains a chemically sound option at all, independent of whether its own top-1 ranking found it.

\subsection{Two selectivity case studies}\label{55-case-studies}

Sections~\ref{51-disconnection-recallk}--\ref{53-pluggability-reranking-arbitrary-backbones-with-chemharness} show \emph{that} competing systems hallucinate; the cases below show \emph{how}. We first examine two targets in detail, comparing every backbone's delivered top-1 route side by side, and then summarize two further cases. In each case the hallucination is not an artifact of one architecture: specialized models and frontier LLMs, trained and prompted independently, converge on the same chemically unexecutable disconnection.

\subsubsection*{Case 1: chemoselectivity (carboxylate vs phenoxide) in a Williamson etherification}
\begin{figure}[H]
  \centering
  \begin{subfigure}{\linewidth}
    \centering
    \includegraphics[width=0.7\linewidth]{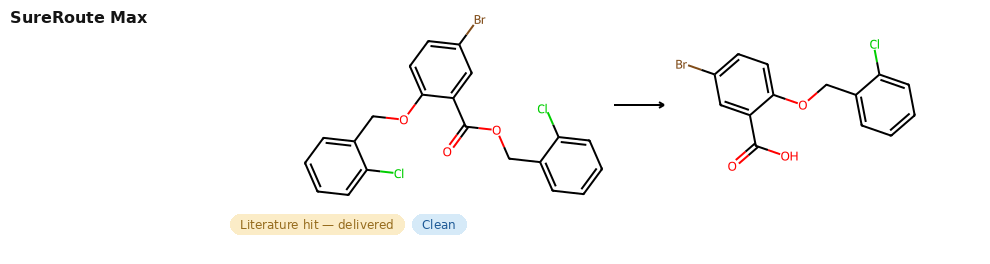}
    \caption{\textbf{SureRoute Max}: literature ester-hydrolysis route.}
    \label{fig:case-t1-sureroute}
  \end{subfigure}\vspace{0.35em}

  \begin{subfigure}{\linewidth}
    \centering
    \includegraphics[width=0.90\linewidth]{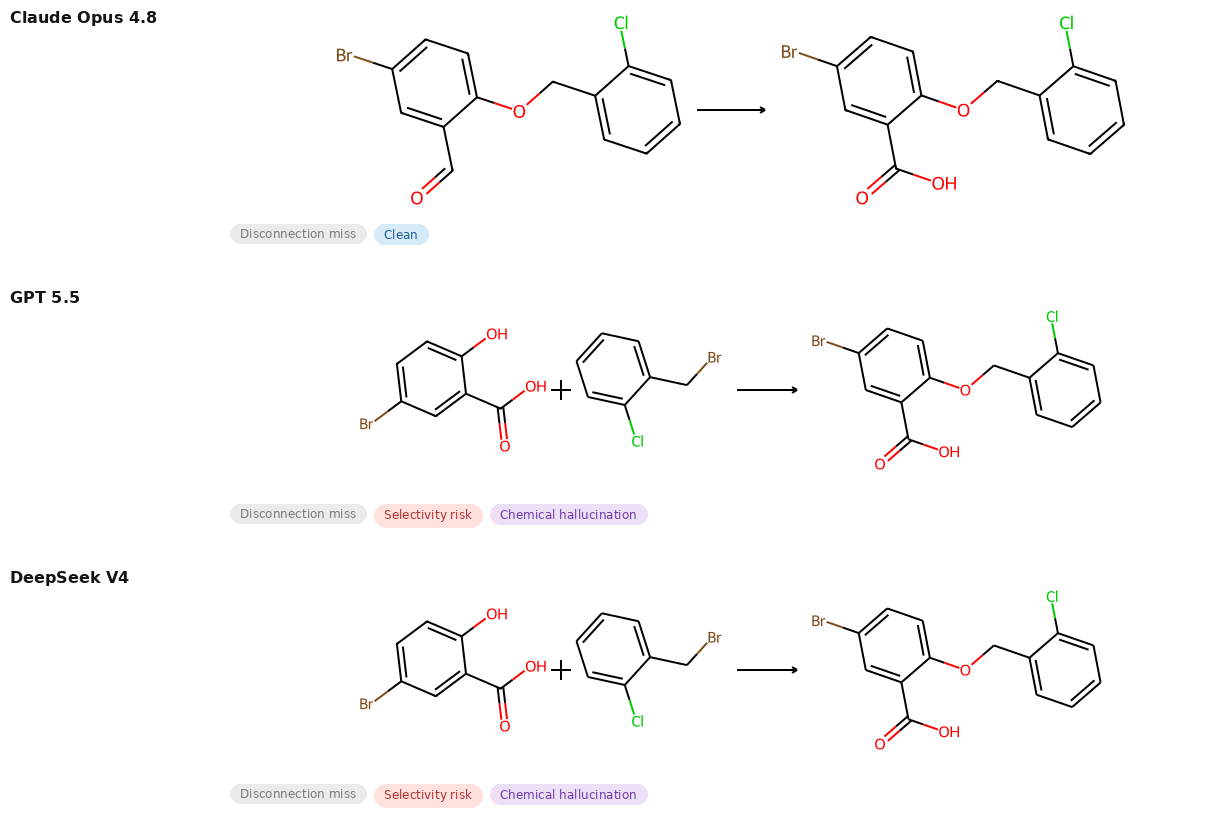}
    \caption{Frontier LLMs: Claude Opus 4.8 gives a clean non-reference oxidation; GPT-5.5 and DeepSeek V4 return the competitive ether disconnection.}
    \label{fig:case-t1-llm}
  \end{subfigure}\vspace{0.35em}

  \begin{subfigure}{\linewidth}
    \centering
    \includegraphics[width=0.90\linewidth]{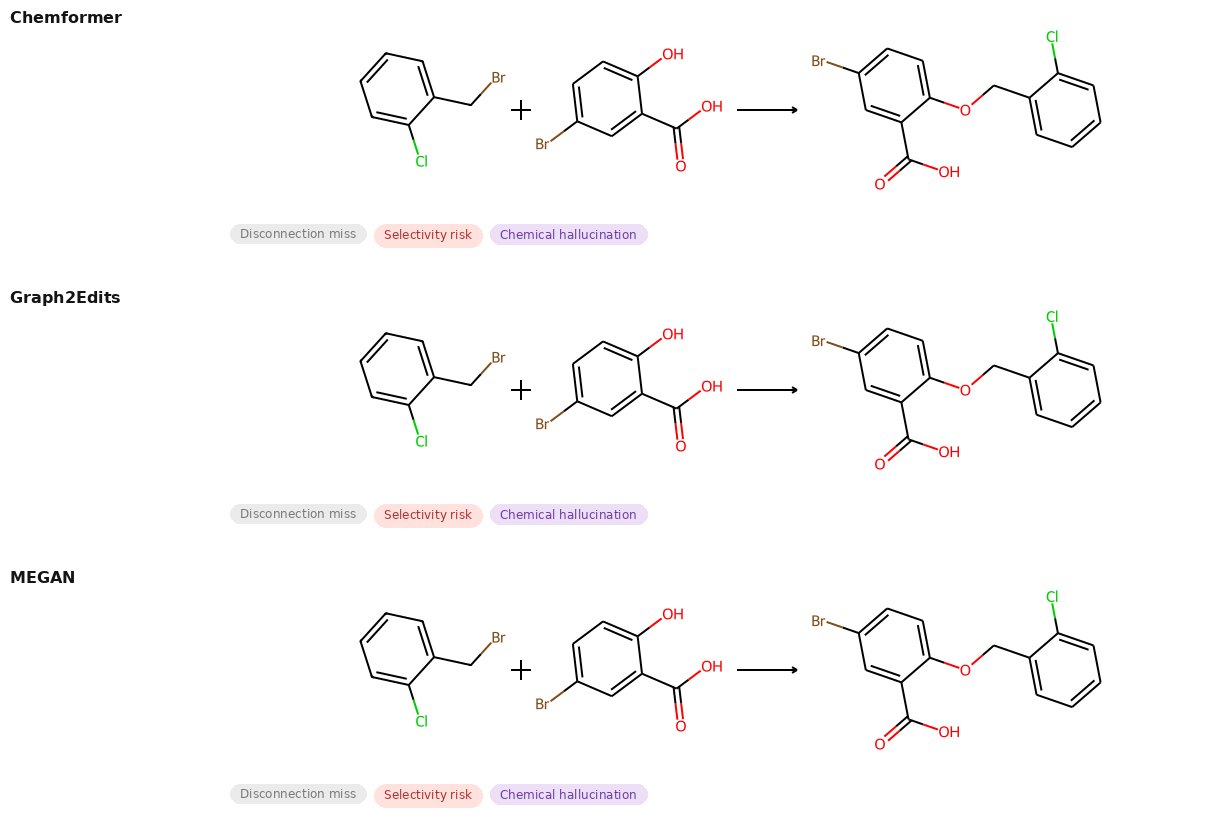}
    \caption{Chemformer, Graph2Edits, and MEGAN all return the same competitive Williamson disconnection.}
    \label{fig:case-t1-sota}
  \end{subfigure}

  \caption{\textbf{Case 1: chemoselectivity (carboxylate vs phenoxide).} Specialized models and two frontier LLMs converge on a chemoselectively wrong Williamson disconnection; SureRoute retains the literature ester-hydrolysis route.}
  \label{fig:case-t1}
\end{figure}

The target carries a free carboxylic acid \emph{ortho} to a benzyl aryl ether. The obvious retrosynthetic move is to cut the ether bond, giving 5-bromosalicylic acid and 2-chlorobenzyl bromide. Atomically the disconnection balances, and it looks like a textbook Williamson ether synthesis. Chemically it does not work as written: the carboxylic acid is substantially more acidic than the phenol, so under the basic conditions the alkylation requires, the carboxylate is deprotonated preferentially and is alkylated first, delivering the benzyl \emph{ester} rather than the desired aryl ether. Resolving this requires either protecting the acid or exploiting a large reactivity difference that the proposed conditions do not supply. The literature route avoids the ambiguity entirely by installing the ether earlier and unmasking the acid last, via a simple ester hydrolysis (Figure~\ref{fig:case-t1}(a)).

Every specialized model we tested proposes the Williamson disconnection as its top-1 (Figure~\ref{fig:case-t1}(c)), and two of the three frontier LLMs do the same (Figure~\ref{fig:case-t1}(b)). ChemHarness flags all of them for O-alkylation competition. Claude Opus 4.8 is the one backbone that avoids the hallucination, proposing an aldehyde oxidation that ChemHarness marks clean --- a chemically executable route, though not the reference one, and therefore still a recall miss under the protocol of Section~\ref{43-evaluation-protocol}. This is precisely the distinction the Chemical Hallucination metric is designed to expose: a disconnection miss and a chemical hallucination are different failures, and only the latter would be unsafe to reinforce in a self-improvement loop.

\subsubsection*{Case 2: regioselectivity --- dichloropyrimidine (\texttt{ACYLSC046})}
The target contains a dichloropyrimidine scaffold with two potentially reactive C–Cl bonds. A natural retrosynthetic proposal couples 2,4-dichloropyrimidine with 3-chlorophenylboronic acid. However, the two chlorinated positions are not equally favored under conventional cross-coupling conditions: coupling is expected to occur preferentially at the more activated position, producing the undesired regioisomer rather than the target (Figure~\ref{fig:case-t2}(a)).
All seven specialized models place this disconnection at top-1, as does Claude Opus 4.8 (Figure~\ref{fig:case-t2}(b)--(c)); GPT 5.5 gives an unmatched answer, and DeepSeek V4 does not return any route. ChemHarness flags each proposal for regioselectivity risk. The selectivity trap was independently annotated by an XtalPi chemist. Here, no backbone recovers the literature disconnection. Fluency and template coverage alike are blind to the selectivity problem, and only the verification layer separates the executable route from the hallucination one.

\begin{figure}[!htbp]
  \centering

  \begin{subfigure}{\linewidth}
    \centering
    \includegraphics[width=\linewidth]{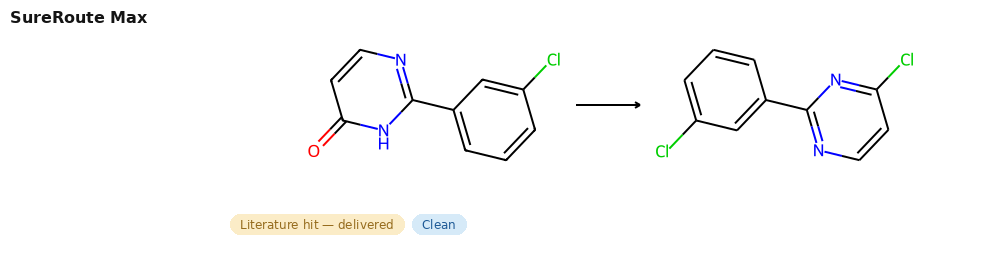}
    \caption{\textbf{SureRoute}: matched with the literature route from the pyrimidinone.}
    \label{fig:case-t2-sureroute}
  \end{subfigure}

  \caption[]{\textbf{Case 2: regioselectivity in a dichloropyrimidine coupling.}
  Continued on the next page.}
\end{figure}

\begin{figure}[!t]
  \ContinuedFloat
  \centering

  \begin{subfigure}{\linewidth}
    \centering
    \includegraphics[width=\linewidth]{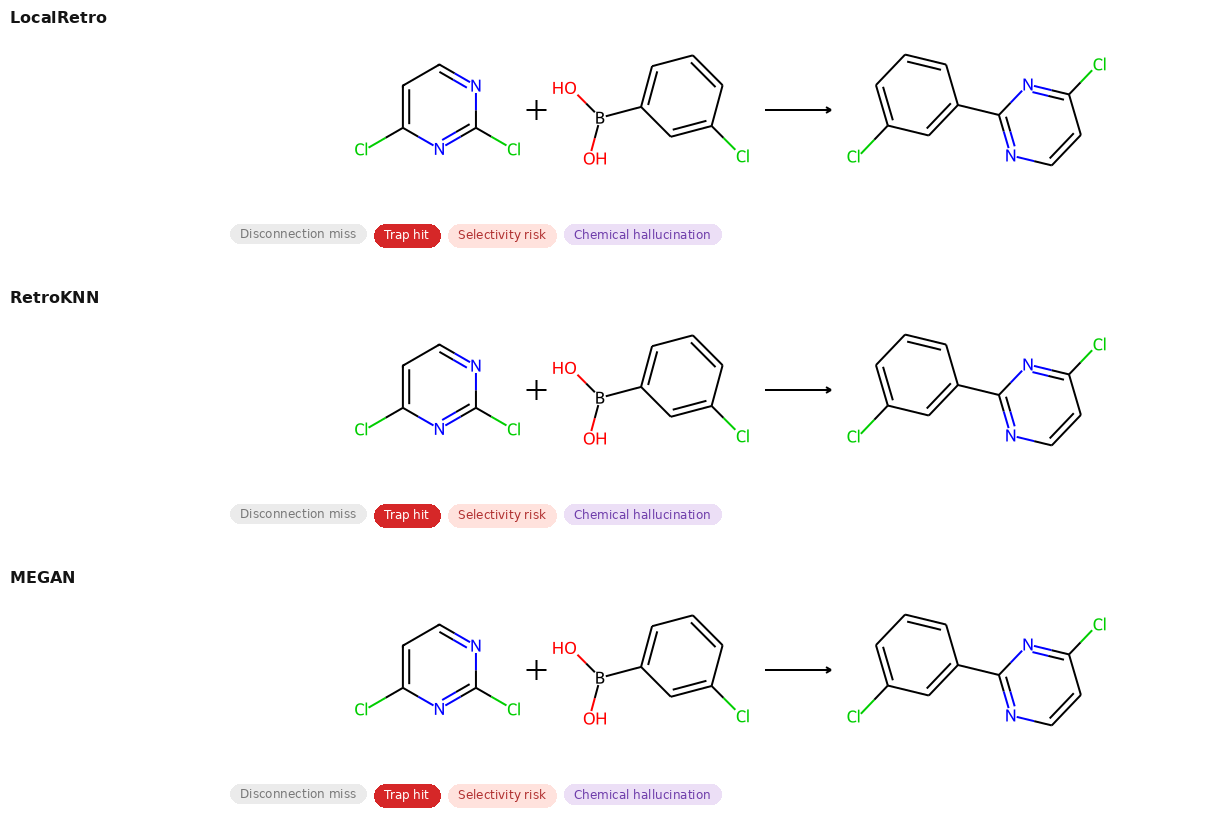}
    \caption{Specialized retrosynthesis models such as LocalRetro, RetroKNN,
    and MEGAN all return the same selectivity-unsafe disconnection.}
    \label{fig:case-t2-sota}
  \end{subfigure}

  \vspace{0.4em}

  \begin{subfigure}{\linewidth}
    \centering
    \includegraphics[width=\linewidth]{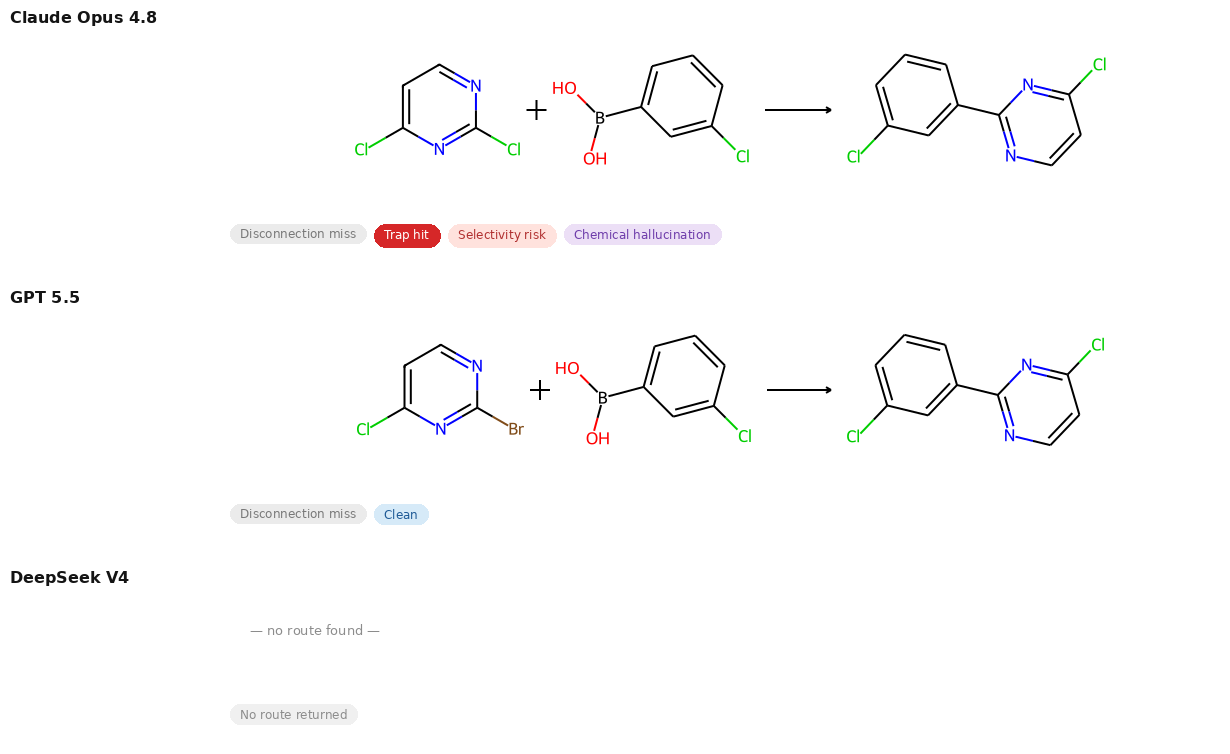}
    \caption{Claude Opus 4.8 proposes the same disconnection without resolving
    the competing C--Cl site.}
    \label{fig:case-t2-llm}
  \end{subfigure}

  \caption{\textbf{Case 2: regioselectivity in a dichloropyrimidine coupling.}
  Competing C--Cl sites make the coupling position ambiguous; SureRoute ranks
  the literature-supported pyrimidinone route first.}
  \label{fig:case-t2}
\end{figure}

\newpage
\subsubsection*{Further Chemical Hallucination cases}

Additional cases follow the same pattern --- all seven single-step models (and, in most cases, one or more LLMs) converge on a single flagged disconnection:

\begin{figure}[H]
  \begin{subfigure}{\linewidth}
    \centering
    \includegraphics[width=0.92\linewidth]{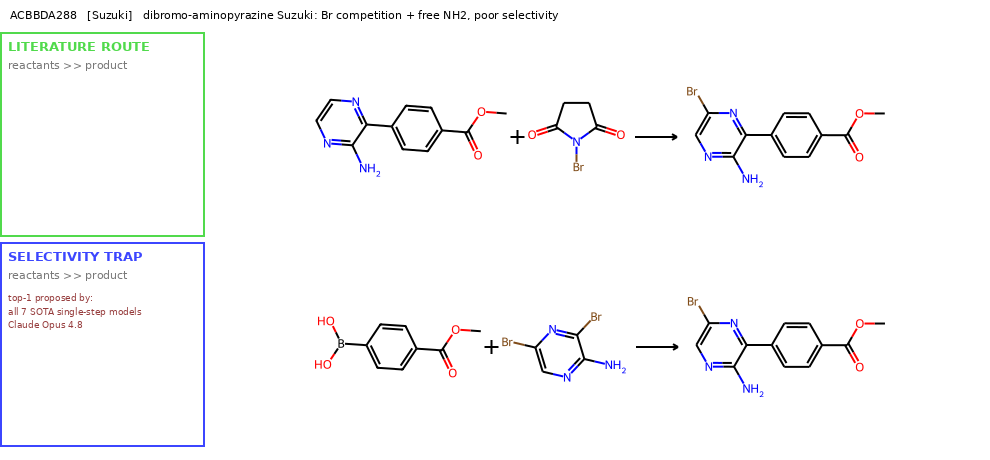}
    \caption{}\label{fig:hallucination-b}
  \end{subfigure}\vspace{0.5em}

  \begin{subfigure}{\linewidth}
    \centering
    \includegraphics[width=0.92\linewidth]{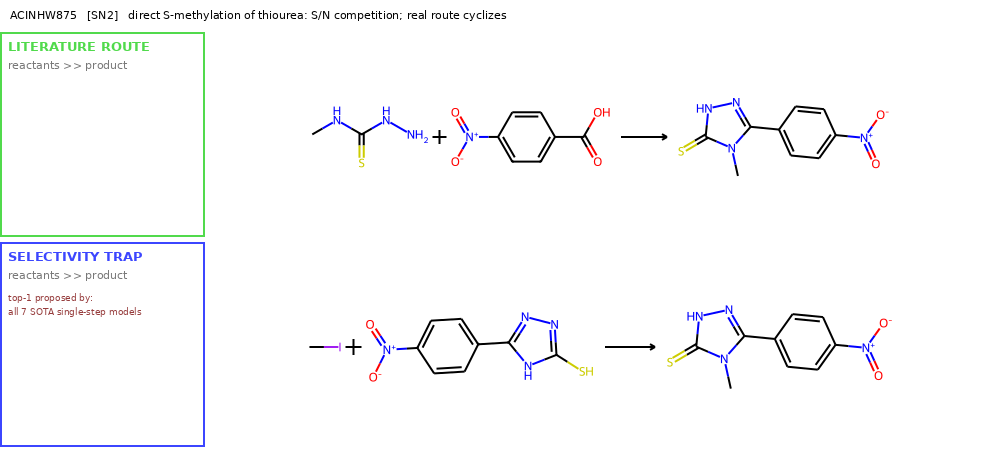}
    \caption{}\label{fig:hallucination-c}
  \end{subfigure}

  \caption{Further selectivity-hallucination cases. Blue boxes show literature routes; grey boxes show atom-mapping consistent but selectivity-unsafe top-1 predictions. ChemHarness screens precisely this failure mode.}
  \label{fig:selectivity-hallucinations}
\end{figure}
\FloatBarrier

\begin{itemize}
\item
  \textbf{Dibromo-aminopyrazine Suzuki coupling} (\texttt{ACBBDA288}): competing bromides plus an unprotected free amine create a selectivity problem; the literature route disconnects at a single, unambiguous aromatic ring position.
\item
  \textbf{Thiourea \emph{S}-methylation} (\texttt{ACINHW875}): competing \emph{S}- vs.\ \emph{N}-methylation leaves the product structure ambiguous; the literature route proceeds through a cyclization step that fixes the reactive site.
\end{itemize}

In each, all seven single-step models' top-1 prediction lands on a chemically hallucination route; only SureRoute's ChemHarness-based reranking anchors the top-1 slot on the literature route (Figure~\ref{fig:selectivity-hallucinations}).

\subsection{Multi-step extension}\label{56-multi-step}

\begin{figure}[!t]
  \centering
  \begin{subfigure}{\linewidth}
    \centering
    \includegraphics[width=0.95\linewidth]{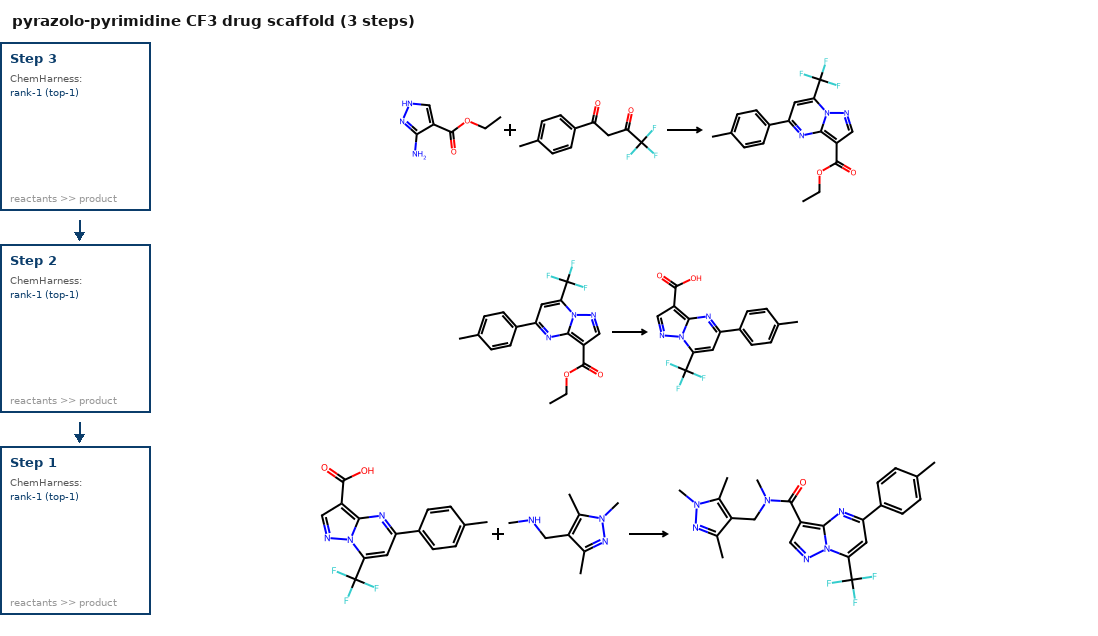}
    \caption{}\label{fig:multistep-a}
  \end{subfigure}\vspace{0.35em}

  \begin{subfigure}{\linewidth}
    \centering
    \includegraphics[width=0.95\linewidth]{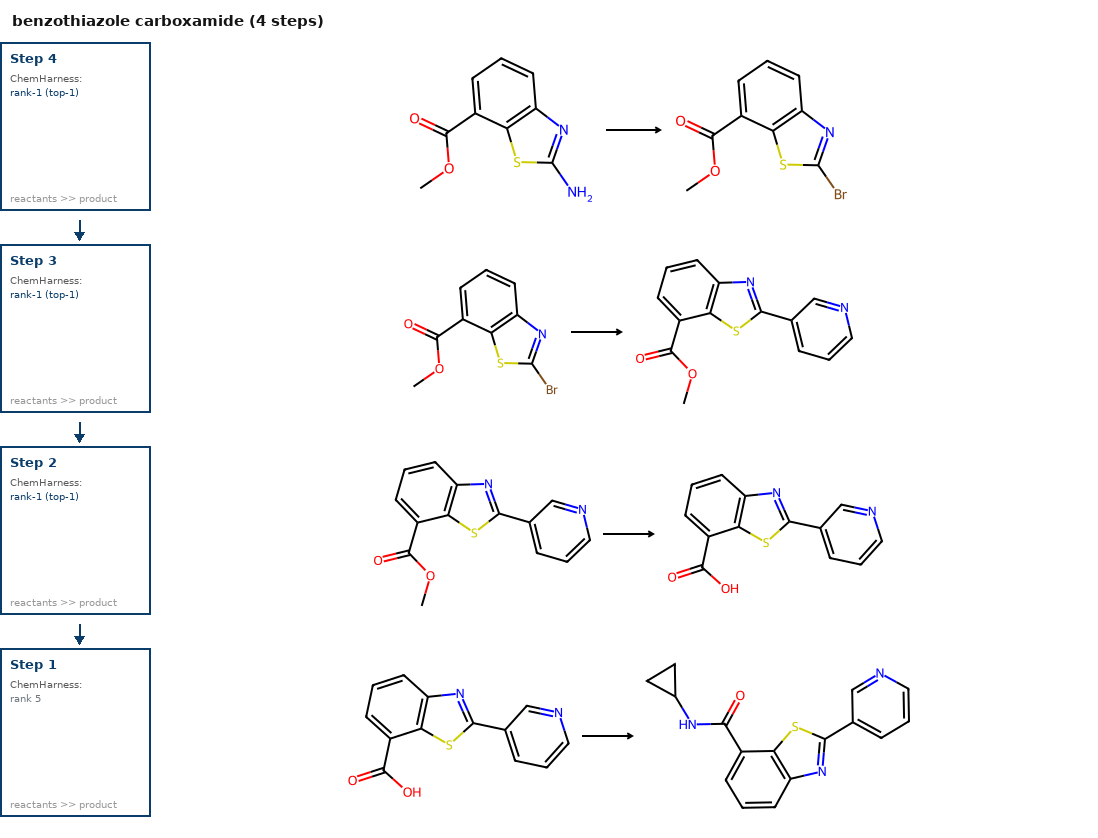}
    \caption{}\label{fig:multistep-b}
  \end{subfigure}

  \caption{Two literature multi-step routes recovered by ChemHarness-guided step-wise prediction. Blue annotations report the rank of the reference disconnection at each step; the correct choice remains top-ranked at nearly every step.}
  \label{fig:multistep}
\end{figure}

While the main evaluation is single-step, we additionally verified that the same reliability advantage compounds over multi-step routes, using a separate set of literature targets (3--5 steps) with fully documented experimental routes. Under step-wise interactive prediction (ChemHarness re-evaluated at each step), we recovered the complete literature route---every step, in the correct order, with the correct disconnection ranked first at nearly every individual step (Figure~\ref{fig:multistep}).   Beyond these benchmarks, hundreds of expert
chemists at XtalPi interact with SureRoute in production, annotating
mis-ranked routes and continually evolving the system.

\FloatBarrier

\section{Conclusion}\label{7-conclusion}

We introduced SureRoute, a chemically verified self-improving retrosynthesis platform that pairs a multi-model disconnection ensemble and literature retrieval with ChemHarness, a rule-based, mechanism-aware reliability layer that explicitly screens for functional-group competition and mechanistic implausibility. On an internal 350-molecule benchmark, SureRoute achieves 2.2--3.5\ensuremath{\times} the top-1 recall of both SOTA single-step models and frontier LLMs, and---more importantly---a Chemical Hallucination rate of 4.6\%, a 4--6\ensuremath{\times} reduction relative to frontier LLMs. We showed that this reliability gain is not tied to our own ensemble: applying ChemHarness reranking to any backbone's raw candidates, including LLMs it was never tuned against, drives ChemHarness-detectable top-1 Chemical Hallucination toward near-zero, establishing ChemHarness as a model-agnostic, pluggable reliability layer rather than a component specific to one system. 

More broadly, we position SureRoute's contribution within a wider trend of self-improving retrosynthesis and scientific-AI systems, where the central open problem has shifted from generation capability to verification capability: a self-improvement loop is only as trustworthy as whatever decides which self-generated outputs to reinforce, and a loop that grades its own output risks compounding its own mistakes rather than correcting them. ChemHarness is designed to occupy the space between wet-lab-grade trust and unit-test-grade cost---an executable, auditable verifier that lets SureRoute close a rule-refinement loop over real production failures without either the expense of physical validation or the risk of self-graded reward. This verifier-anchored, compounding advantage is, we argue, difficult for a general-purpose LLM to replicate through scale or prompting alone, because it depends on continually accumulating and correcting against real deployment failures rather than on any fixed amount of pretraining. 

\newpage
\bibliographystyle{plainnat}
\bibliography{references}

\clearpage

\appendix
\section{LLM Evaluation Protocol}
\label{appendix-a-llm-evaluation-protocol}

We evaluate frontier large language models for retrosynthesis using a unified
prompt template. All generated candidates are processed using the same
downstream standardization and matching pipeline as specialized retrosynthesis
models (Section~\ref{43-evaluation-protocol}). The full prompt consists of a system instruction and a
user message containing two in-context demonstrations followed by the query.

\subsection{Prompt Template}

\begin{promptbox}{System}
You are an expert organic chemist specializing in single-step retrosynthesis.
Given a target molecule, propose reactant sets that could be combined in one
reaction step to synthesize it.
\end{promptbox}

\begin{promptbox}{User}
(*@\textbf{\sffamily Demonstration 1 (amide coupling)}@*)
Target:
O=C(Nc1ccccc1)c1ccccc1
Answer:
O=C(O)c1ccccc1.Nc1ccccc1
O=C(Cl)c1ccccc1.Nc1ccccc1

(*@\textbf{\sffamily Demonstration 2 (Suzuki coupling)}@*)
Target:
CCOC(=O)c1ccc(-c2ccccc2)cc1
Answer:
CCOC(=O)c1ccc(Br)cc1.OB(O)c1ccccc1
CCOC(=O)c1ccc(Br)cc1.CC1(C)OB(c2ccccc2)OC1(C)C

(*@\textbf{\sffamily Query}@*)
Propose up to 5 distinct single-step retrosynthetic disconnections for the
target below.

Rules:
- Each line represents one reactant set.
- Reactants are separated by ".".
- Order candidates from most to least likely.
- Output only reactant SMILES.
- Do not provide explanations or reaction arrows.
- Reactants should represent purchasable-style building blocks.

Target:
{target SMILES}
Answer:
\end{promptbox}

This prompt elicits ranked candidate lists rather than a single prediction,
enabling direct comparison between LLMs and specialized retrosynthesis models
under the same recall@$K$ evaluation protocol.

\clearpage
\section{Target Visualization}
\label{sec:target-visualization}
\vspace{-0.5em}
\begin{center}
  \includegraphics[
    height=0.865\textheight,
    keepaspectratio
  ]{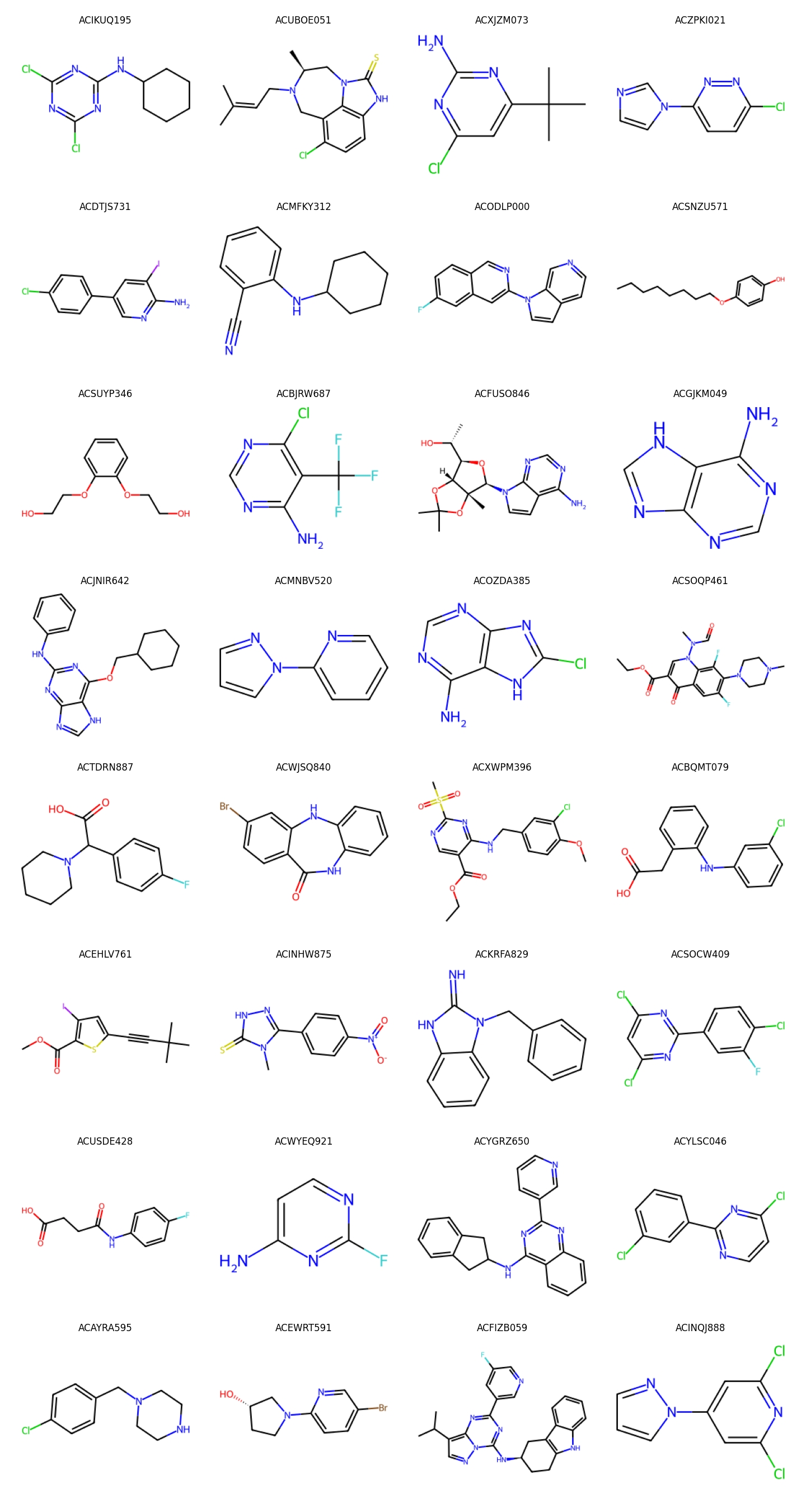}
  \vspace{-0.35em}
  \captionof{figure}{\textbf{Target visualization of the industrial retrosynthesis evaluation set.}}
  \label{fig:target-visualization}
\end{center}

\end{document}